# Element Abundances of Solar Energetic Particles and the Photosphere, the Corona, and the Solar Wind

**Donald V. Reames**

Institute for Physical Science and Technology, University of Maryland, College Park, MD 20742–2431 USA**;** dvreames@umd.edu

**Abstract:** From a turbulent history, the study of the abundances of elements in solar energetic particles (SEPs) has grown into an extensive field that probes the solar corona and physical processes of SEP acceleration and transport. Underlying SEPs are the abundances of the solar corona, which differ from photospheric abundances as a function of the first ionization potentials (FIPs) of the elements. The FIP-dependence of SEPs also differs from that of the solar wind; each has a different magnetic environment, where low-FIP ions and high-FIP neutral atoms rise toward the corona. Two major sources generate SEPs: The small "impulsive" SEP events are associated with magnetic reconnection in solar jets that produce 1000-fold enhancements from H to Pb as a function of mass-to-charge ratio $A/Q$, and also 1000-fold enhancements in $^3$He/$^4$He that are produced by resonant wave-particle interactions. In large "gradual" events, SEPs are accelerated at shock waves that are driven out from the Sun by wide, fast coronal mass ejections (CMEs). $A/Q$ dependence of ion transport allows us to estimate $Q$ and hence the source plasma temperature $T$. Weaker shock waves favor the reacceleration of suprathermal ions accumulated from earlier impulsive SEP events, along with protons from the ambient plasma. In strong shocks, the ambient plasma dominates. Ions from impulsive sources have $T \approx 3$ MK; those from ambient coronal plasma have $T = 1 - 2$ MK. These FIP- and $A/Q$-dependences explore complex new interactions in the corona and in SEP sources.

## 1. Introduction

Solar energetic particles (SEPs) carry an invisible, yet direct, record of high-energy physics at the Sun that we have slowly learned to read in a field that is dominated by beautiful images of the Sun. Improving measurements of the relative abundances of the chemical elements in SEPs have been shown to carry an imprint of corresponding abundances in the solar corona, of the source conditions, and of characteristic modifications that show unique histories of SEP acceleration and transport. This is becoming a rich field. However, the path to the identification of the sources of acceleration has not been tranquil [1].

*1.1.A Turbulent History: Gradual and Impulsive SEP Events*

Forbush first observed the effects of SEPs [2], as increases in GeV particles that caused nuclear cascades through the atmosphere to produce what we now call ground-level events (GLEs). These SEPs were visually associated with solar flares and it was commonly assumed that the flares themselves caused SEP events in some way, but there were problems; for example, SEPs that were associated with these flares were eventually found to span more than 180° in solar longitude, which suggests a broad spatial distribution. How could the SEPs from a point-source flare cross magnetic field lines over such a distance? Newkirk and Wenzel [3] proposed the "birdcage" model where SEPs could follow coronal magnetic loops that then spread across the Sun like the wires of a birdcage. The birdcage model reigned many years, but Mason, et al. [4] argued that abundances of ions of different rigidity were unaffected by a trip through such a complex magnetic birdcage; they must arise from large-scale shock acceleration. At the same time, Kahler et al. [5] found a 96% correlation between large energetic SEP events and wide, fast coronal mass ejections (CMEs), which had just been discovered several years before. We now know that CMEs drive extensive expanding sun-spanning shock waves, where SEPs are accelerated to produce what we now call "long-duration" or "gradual" SEP events. Considerable controversy arose again with the publication of Gosling's [6] review article entitled "The Solar Flare Myth", which was alleged to "wage an assault on the last 30 years of solar-flare research" [7], apparently based upon a fear that flare research would no longer be funded unless flares produced the radiation that was hazardous to astronauts, which they actually do not. Of course, flare science remains quite healthy 26 years later.

However, improving measurements made the shock acceleration of gradual SEPs increasingly difficult to ignore [1,8–13]. These included onset timing of SEPs [14,15], multi-spacecraft studies [16] with improved shock imaging [17–22], temperature and ionization-state measurements [13,23–28], shock modeling [24–35], and correlations of SEP intensities with CME speed [22,36], which continued to strengthen the evidence [1,34].

Meanwhile, there was early evidence for a completely different kind of SEP event being first noted by the abundance of the isotope $^3$He, which was enhanced by orders of magnitude. The ratio $^3$He/$^4$He ≈ 5×10$^{-4}$ in the solar wind, but Serlemitsos and Balasubrahmanyan [37] found $^3$He/$^4$He = 1.52 ± 0.10 in a small SEP event, completely unaccompanied by any $^2$H, $^3$H, Li, Be, or B (Be/O and B/O < 4×10$^{-4}$), failing to suggest any evidence of nuclear fragmentation at all. These $^3$He-rich events also had enhancements of Fe/O and they were associated with streaming electrons [38] and the type III radio bursts [39] that they produce. Current theory [40] suggests that $^3$He resonates with electromagnetic ion cyclotron waves that were generated by the streaming electrons. More recently, the enhancements of heavy elements have been found to extend across the periodic table from H and He to Pb with a 1000-fold enhancement in the elements (76 ≤ Z ≤ 82)/O, relative to that in the gradual SEP events or the solar corona [41–44]. These heavy-element enhancements are found to originate in magnetic reconnection [45] on open field lines in solar jets [46–48]. Energetic ions from similar acceleration in flares are magnetically trapped on loops and they deposit their energy in the footpoints producing $\gamma$-ray lines [49–51] and hot (10–40 MK), bright flares, but the charged ions do not escape.

The time scale of associated X-rays, which now seem less clear and considerably less relevant, originally distinguished impulsive and gradual SEP events. It is true that most impulsive events last for hours, while most gradual events last days, but these times are not well resolved. It now seems that the cleanest separation of the event types comes from bimodal enhancements in Fe/O. Impulsive SEP events, which are associated with solar jets, have over four times the average Fe/O of gradual events [13,44].

*1.2. SEP Abundance Measuements*

To evaluate element abundance measurements of SEPs, it is important to understand the relative simplicity of those measurements. Logically, the measurements in the most common instruments consist of a thin detector measuring the energy-loss rate *dE/dx*, followed by a thick detector *E*, in which the ion stops, insured by an anti-coincidence detector. Si solid-state detectors are used in recent instruments. Figure 1 shows response "matrices" for two instruments [1] (Chapt. 7).

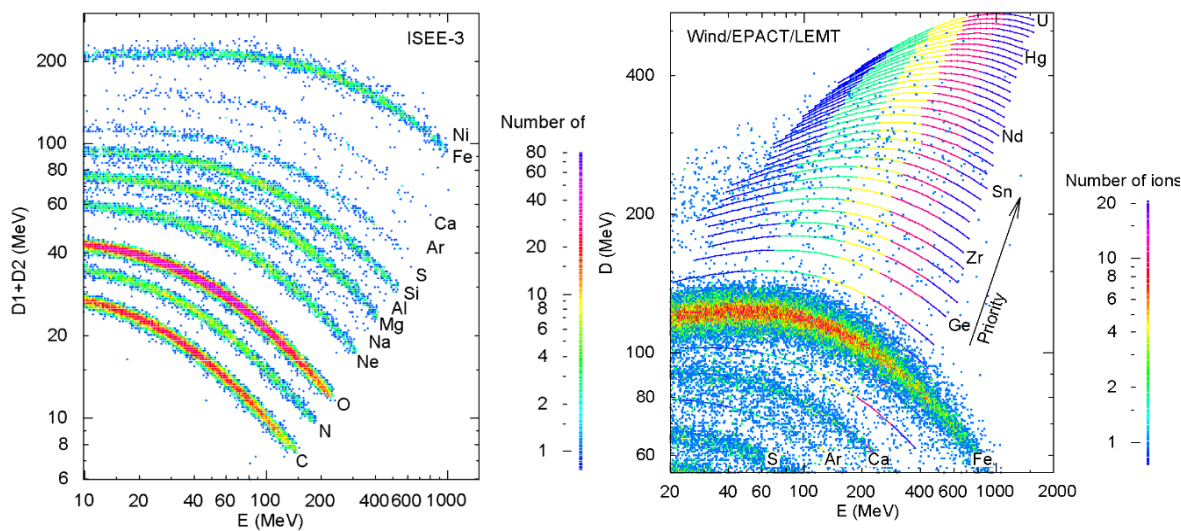

**Figure 1.** Response "matrices", showing the number of ions for each *dE vs. E* bin, separate ions into tracks of increasing energy for each element as listed. Light elements are well resolved [52] in the *left panel* where energy deposit in thin detectors D1 and D2, are plotted against energy deposit in the thick E detector where ions stop. The *right panel* shows similar D *vs.* E plot for a telescope with dynamic range above Fe to Hg and U, where heavy elements above Fe are only grouped [41,42,53].

Unlike other abundance measurements, such as spectral lines or the solar wind, SEPs measure a broad range of elements at once and all ionization states are summed. Abundances are simply obtained by counting ions in well-defined intervals of particle velocity, or MeV amu$^{-1}$, during appropriate time intervals. Relative abundances of C, N, and O, or Ne, Mg, Si, and Fe are already apparent from these plots of raw data. High- and low-FIP elements, like Ne and Mg, are treated equally in SEPs. These instruments measure ions above about 1 MeV amu$^{-1}$. A similar quality measurements can be obtained below 1 MeV amu$^{-1}$ while using time-of-flight *vs. E* techniques [54] and instruments with position-sensing detectors that correct for ion arrival directions can resolve isotopes up to Fe [55,56] at >10 MeV amu$^{-1}$.

## 2. Abundances: Separating Dependence upon FIP and *A/Q*

From the earliest measurements of element abundances in SEP events, there were attempts to make comparisons with measurements of abundances in the solar photosphere, which were also evolving with time, and with abundances of galactic cosmic-ray sources [57]. It was clear that the abundances, averaged over many large SEP events and divided by the corresponding photospheric abundances, had a dependence upon the first ionization potential (FIP) of the elements. It was understood that high-FIP (> 10 eV) elements (*e.g.* He, C, O, Ne) were neutral atoms in the photosphere and, as they began to cross the chromosphere, but the low-FIP elements (*e.g.* Mg, Si, Fe) were ions that could be affected by electromagnetic fields. All of the elements become highly ionized in the ≈1 MK corona. Figure 2 shows a modern version of the "FIP effect".

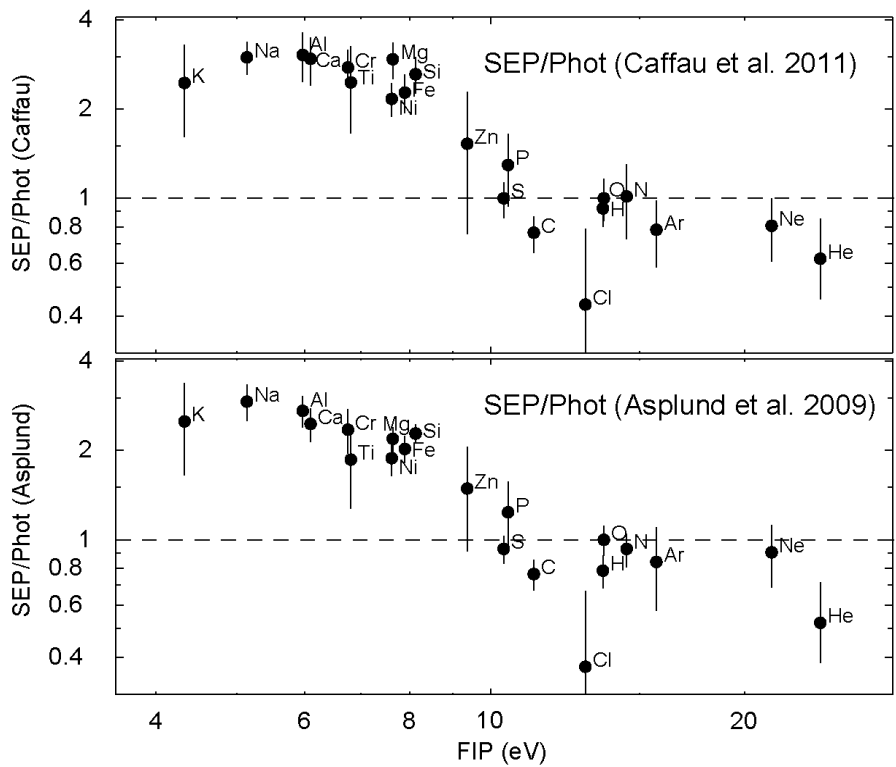

**Figure 2.** Average solar energetic particle (SEP) abundances [52,58,59] divided by corresponding photospheric abundances are plotted *vs.* first ionization potential (FIP) comparing photospheric abundances of Asplund *et al.* [60] (*lower panel*) and Caffau *et al.* [61,62] (*upper panel*). See Appendix for numerical values.

In a classic review, Meyer [63] found that SEP-event abundances had a common FIP effect, but individual events also showed a dependence on mass-to-charge ratio *A*/*Q*; this was most clearly observed as a power-law by Breneman and Stone [64]. We understand this *A*/*Q*-dependence as mainly a result of scattering of ions by magnetic irregularities during transport away from a shock-acceleration source. Scattering depends approximately as a power law upon the magnetic rigidity of the ion, $M_u(A/Q)\beta\gamma$, where the mass unit $M_u$ = 931 MeV, $\beta = v/c$, $\gamma = (1-\beta^2)^{-1/2}$, and $v$ is the ion velocity. These ratios depend upon *A*/*Q* since abundance ratios ions are compared at the same velocity. Thus, for example, Fe/O will be enhanced early in an SEP event but depleted later since Fe scatters less than O; this effect will be spread in longitude by solar rotation [12,13,59]. Thus, averaging over different times in many different gradual SEP events is assumed to recover the abundances at the source. A confirmation of the validity of this assumption is that ions, such as Mg, Si, Fe, and Ni with large differences in *A*/*Q*, but similar values of FIP, have similar enhancements in Figure 2. Power-law dependence of enhancement, relative to these average-SEP or SEP-coronal abundances, *vs.* *A*/*Q* is shown for several situations in Figure 3. The dependences regarding FIP and *A*/Q collaborate; if the FIP-dependence were incorrect, the power-laws could be systematically disrupted, and conversely.

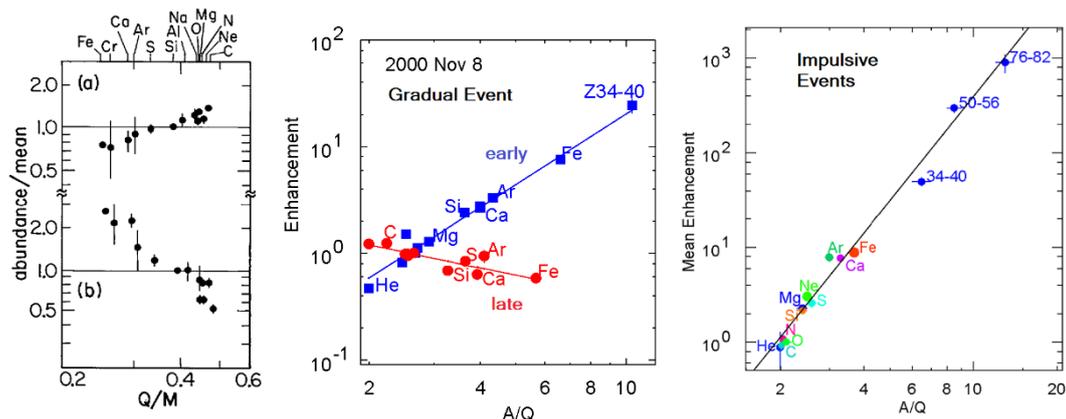

**Figure 3.** Abundance enhancements, relative to coronal abundances, are shown *vs. Q/A* for two gradual events by Breneman and Stone [64] (*left panel*). Variation within an event of enhancement *vs. A/Q* is shown during a gradual event [28] (*center panel*). The average enhancement *vs. A/Q* is shown for impulsive events [44] (*right panel*). Numbers in panels indicate a range of *Z*. Power-law enhancements *vs. A/Q* are quite common in SEP events.

## 3. Source Temperatures and Power Laws in *A*/*Q*

SEP abundances are usually measured without reference to ionization states, but to study power laws in *A*/*Q*, we must know *Q*. Early observations using instruments that did measure SEP ionization states by Luhn et al. [23,24] found the average *Q* of Fe to be 14.1±0.2 in gradual events and 20.5±1.2 in impulsive events, but the latter was subsequently found to depend upon energy and was thus elevated by stripping after acceleration in impulsive SEP events [65]. Breneman and Stone [64] used these early measurements, but such measurements are not available for most SEP events.

However, in the study of abundances in impulsive events [66], it was noted that $^4$He, C, N, and O were all relatively unenhanced, Ne, Mg, and Si were enhanced by a factor of ≈2.5, and Fe was enhanced by a factor of ≈7. This pattern might occur at a temperature range 3 < *T* < 5 MK, where the first group of elements were fully ionized with *Q* = *Z* and *Q*/*A* = 0.5, the second were in a state with two orbital electrons with *Q*/*A* ≈ 0.43, and Fe had *Q*/*A* ≈ 0.28. At higher temperatures, Ne would become stripped; at lower ones, O would capture electrons. With a power-law relationship, the pattern of abundance enhancements directly corresponds to the pattern of *A*/*Q* and, hence, the source plasma temperature.

Figure 4 shows the theoretical relationship between *A*/*Q* and *T*. Improved abundances [44] allowed for the region of impulsive SEP events to be narrowed to 2.5 ≤ *T* ≤ 3.2 MK [66,67] shaded pink and magnified in the figure. In this region, we find *A*/*Q* with the unusual reversed ordering, Ne > Mg > Si, for example, and the newer, more-accurate abundance enhancements in impulsive SEP events were found to behave likewise, as can be seen for the sum of impulsive SEP events in the right-hand panel of Figure 3. Additionally, this power law continues to high *Z* at *T* ≈ 3 MK as a 3.64 ± 0.15 power of *A*/*Q*, on average.

As we proceed to temperatures below 3 MK, as seen in most gradual events, Figure 4 shows that O, N, and then C will become enhanced relative to He, joining a group with Ne. Additionally, with decreasing temperature, Figure 4 shows that S, Si, and then Mg will rise to a group with Ar and Ca. These groups involve filled shells with two or eight electrons and at each *T* the log of the abundance enhancement is linearly related to log *A*/*Q*. Best fits of the enhancements *vs. A/Q* at each temperature allow for us to select the best temperature for impulsive events [66,67] and for gradual events [28], as described in review articles [13] and even a textbook [1].

However, we should note that source abundances of He do show some unique variations in both gradual [68,69] and impulsive [70] events that are not described here. These variations may be caused by unusually slow ionization of He during FIP processing because of its uniquely high FIP of 24.6 eV [71]. Perhaps some jets may suddenly incorporate He, only part of which has been ionized, from chromospheric material, for example.

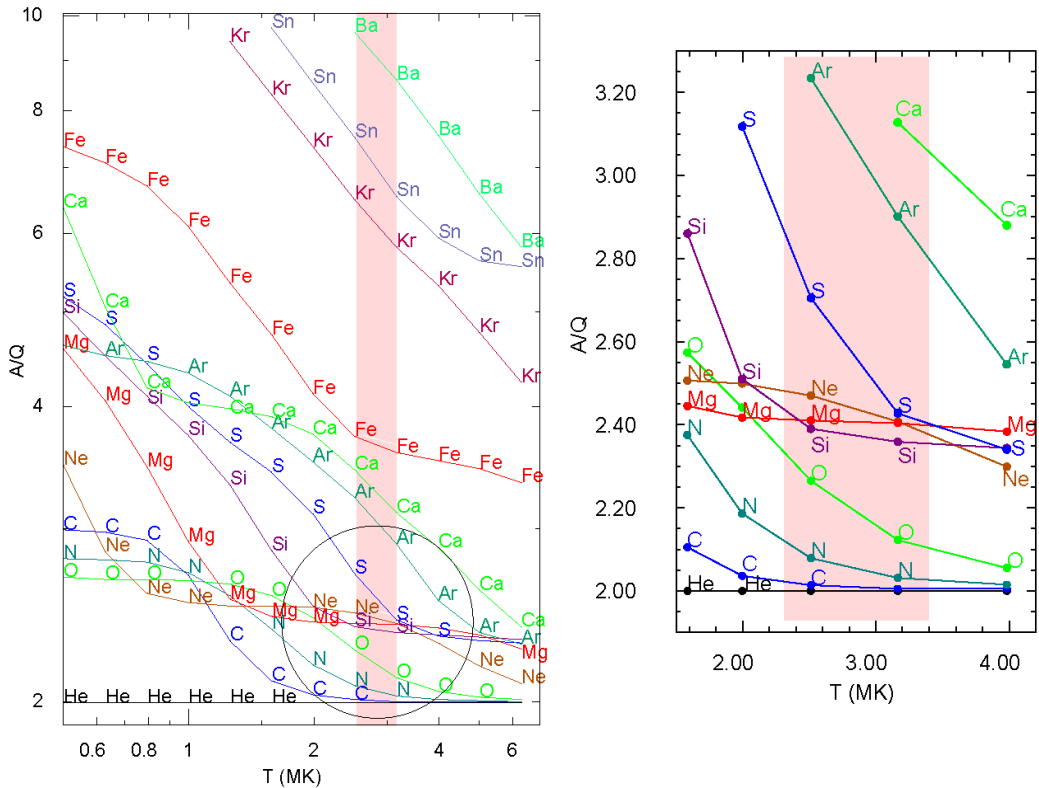

**Figure 4.** The dependence of *A*/*Q* on *T* for selected elements as indicated. Values from He through Fe are from Mazzotta et al. [72]; those for higher Z are from Post et al. [73]. The region circled in the *left panel* is magnified in the *right panel* especially to show the ordering of Ne, Mg, Si, and S in the 2.5–3.2 MK region. As measurements improved, the ordering of abundance enhancements was found to follow the ordering of *A*/*Q* at this source *T* [67,68].

For impulsive SEP events, the power-law fit for elements with 6 ≤ *Z* ≤ 56 and *A*/*Q* ≥ 2 can be projected down to protons at *A*/*Q* = 1 [74]. For the smaller events, which probably only involve magnetic reconnection in solar jets, the agreement of protons is very good, but for larger jets that also produce CMEs capable of driving shock waves fast enough to reaccelerate the SEPs, there can be a proton excess of an order of magnitude or more. Presumably, these shock waves in larger impulsive events sample protons from the ambient plasma in addition to favoring the high intensities of pre-accelerated SEPs that are already enhanced, as shown in Figure 5.

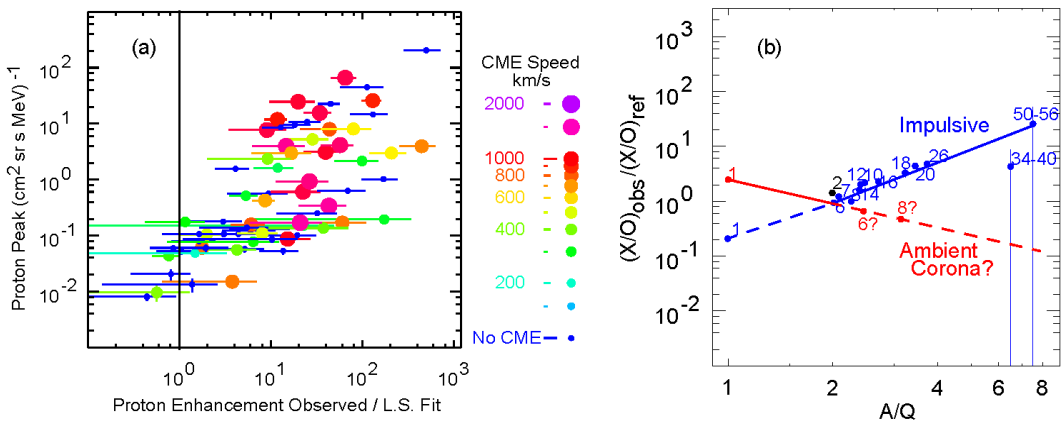

**Figure 5.** Proton excesses for impulsive SEP events and their possible origin: **(a)** Peak intensity *vs.* proton excess shows large excesses for more intense SEP events produced when fast coronal mass ejections (CMEs) can drive shocks. **(b)** Shocks preferentially reaccelerate pre-enhanced impulsive SEP

ions, plus additional protons from the ambient coronal plasma to produce the observed proton excesses [74,75].

Element abundances vary with time in gradual SEP events, so we analyze them in eight-hour intervals. For each interval we study a wide range of possible source temperatures. Each *T* defines *A*/*Q* values and we fit abundance enhancements *vs.* *A*/*Q*, obtaining a best-fit power law and $\chi^2$ for each *T*. We choose the fit and the *T* with the smallest $\chi^2$. Such fits are shown in Figure 6 during two large gradual SEP events in October 2003 [28,75,76]. The best-fit power laws for $Z \geq 6$ are extended down to protons in Figure 6(e).

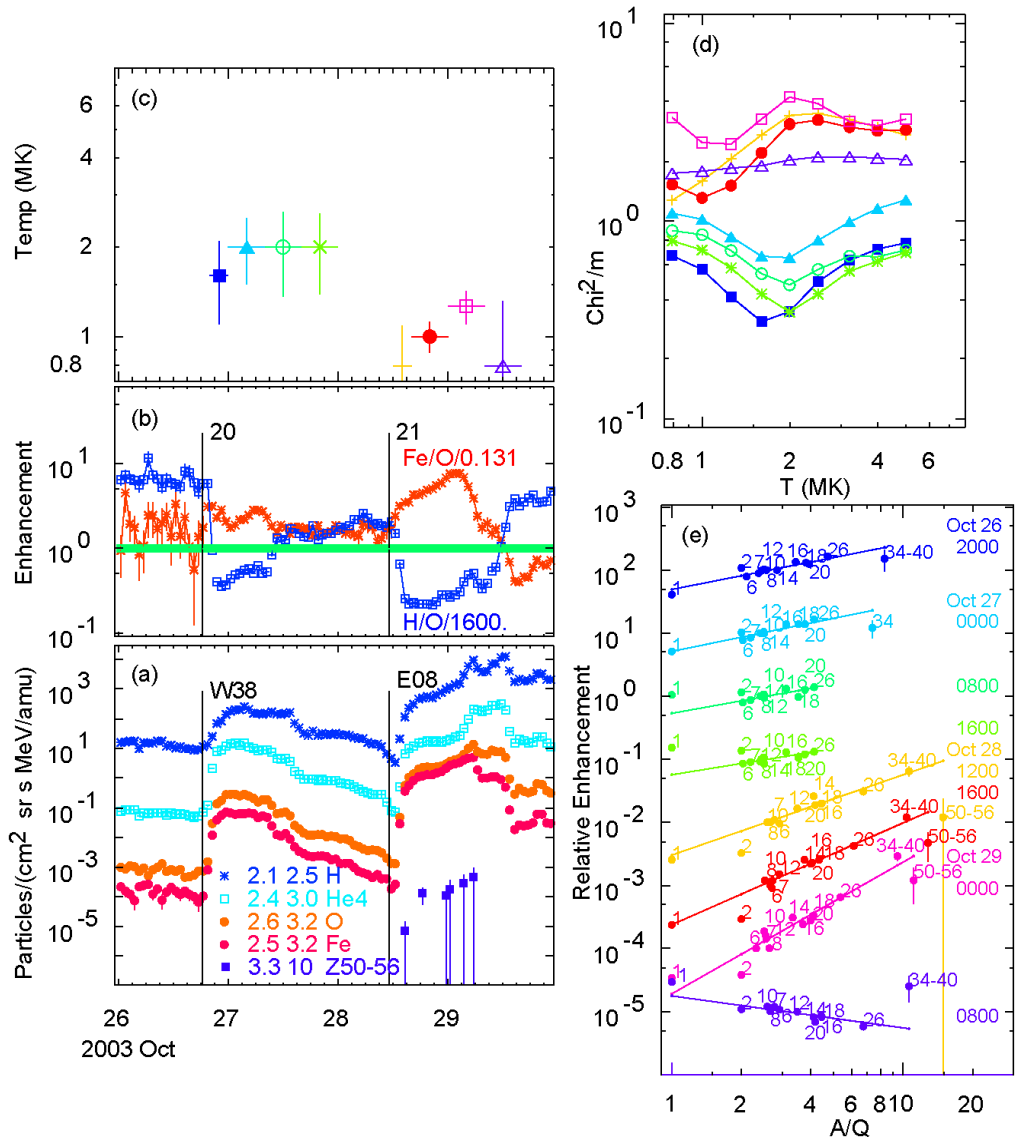

**Figure 6.** Analysis of the 26 and 28 October 2003 SEP events: **(a)** Intensities of H, He, O, Fe, and $50 \leq Z \leq 56$ ions *vs.* time. **(b)** Normalized abundance enhancements H/O and Fe/O *vs.* time. **(c)** Derived source temperatures *vs.* time. **(d)** $\chi^2/m$ *vs.* *T* for each 8-hr interval. **(e)** Enhancements, labeled by *Z*, *vs.* *A*/*Q* for each eight-hr interval shifted ×0.1, with best-fit power law for elements with $Z \geq 6$ extrapolated down to H at *A*/*Q* = 1. Colors correspond for the eight time intervals in (c), (d), and (e) and symbols in (c) and (d); times are also listed in (e). Event onsets are flagged with solar longitude in (a) and event number from [28,76] in (b). The 28 October event produced a ground-level event (GLE).

Note in panel (d) of Figure 6 that the probable source temperature and fit is selected by the minimum in $\chi^2$. When the dependence upon *A*/*Q* is flat, *i.e.* the abundances are nearly coronal, *T* is poorly determined and $\chi^2$ vs. $T$ is flat. In these two events, the proton excesses are minimal and the proton enhancements follow the power-law trend, despite large changes in slope with time.

Approximately 60% of gradual events show no significant proton excesses. These tend to be the largest events with fast, wide CMEs. About 25% of gradual events have been found to have ≈3 MK source temperatures, relatively weak shocks, and abundances with enhancements that look like reaccelerated impulsive SEP events. These events are believed to involve preferential shock acceleration of residual impulsive suprathermal ions [28,77–79]. Reacceleration of residual impulsive suprathermal ions might be especially favored when the magnetic field lies near the plane of the accelerating shock, so only fast particles can overtake the shock from downstream [77,78]. ³He-rich Fe-rich suprathermal ions are commonly observed during brief quiet periods on the active Sun [80–83]; these ions become the seed population for shocks.

These gradual SEP events involving reaccelerated ≈3 MK ions have much smaller variations of source abundance than do single impulsive SEP events of the same temperature, especially in abundance ratios, like He/C, where both ions should be fully ionized at this temperature. Figure 7 shows a suggested explanation for this [75,79].

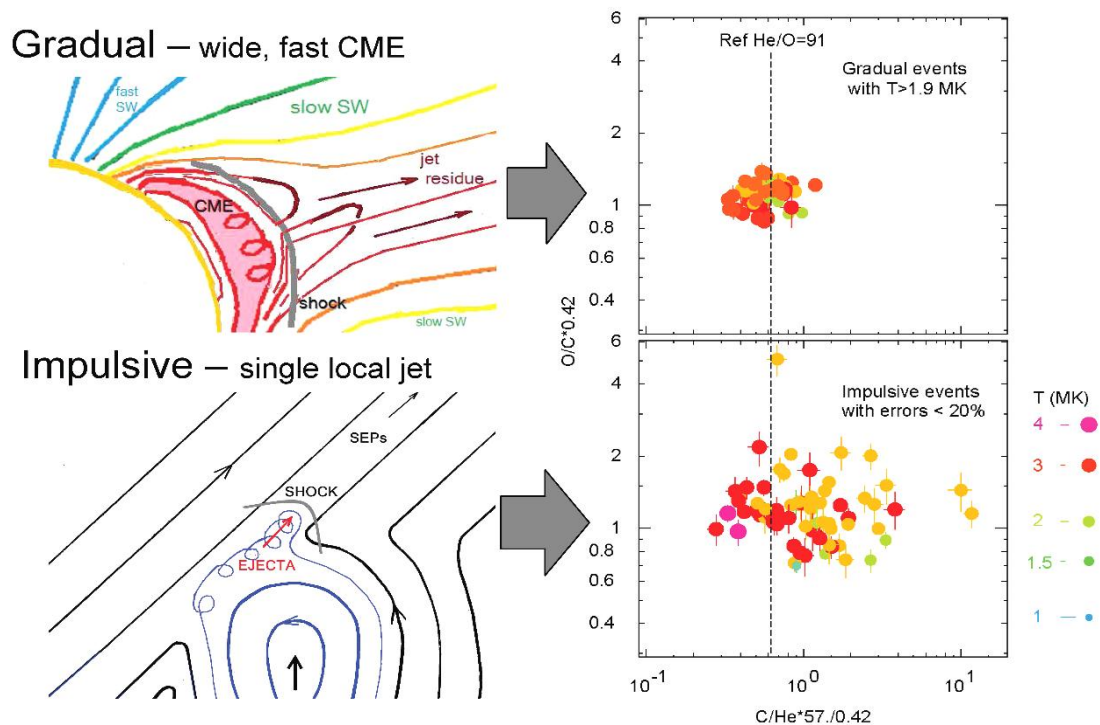

**Figure 7.** Suggested explanations for (*below*) the broad distribution of source abundances in impulsive events from variations in local plasma in individual jet events where ejecta may or may not drive a shock, and (*above*) the narrow distribution in high-*T* gradual events where pre-accelerated impulsive seed populations from many individual jets are averaged by a large shock [75,79].

He/C, both elements with *A*/*Q* = 2 at *T* = 3 MK, and, to a lesser extent, O/C, should be unaltered by acceleration or transport, so these ratios should represent abundances in the source plasma. Each impulsive event samples a small local region, while the gradual events sweep up pre-enhanced suprathermal ions that have been contributed over a large region by many small jets. Many of these jets are too small to be resolvable as individual impulsive SEP events; the number of jets may greatly increase as their size decreases, as is observed for flares where Parker has suggested [84] that numerous nanoflares may even heat the corona. Nanojets may provide the seeds for weak shocks.

However, the really strong shocks in the largest gradual SEP events seem capable of directly sweeping ions from the ambient coronal plasma and the power law in *A*/*Q* extends to include protons. Impulsive suprathermal ions may be present in the seed population, but they are negligible in comparison to the ions of the ambient plasma that are accelerated. Thus, the largest ~60% of gradual SEP events are sufficiently powerful in that pre-accelerated ions are unnecessary [75,76].

## 4. Comparing SEPs with the Solar Wind

The solar wind also samples element abundances from the solar corona and exhibits FIP processing that we can compare with that of the SEPs. Fast (> 500 km s$^{-1}$) solar-wind streams emerge from coronal holes and the slow solar wind (SSW) from interstream regions. The average abundances of elements in SEPs are compared with those of the SSW reviewed by Bochsler [85] in Figure 8 [59,13]. The direct ratio SEP/SSW is shown *vs.* FIP in the lower panel and the abundances, relative to the photosphere, *vs.* FIP are compared in the upper panel. The transition from low to high FIP occurs at a lower value for SEPs than for the SSW. In particular, the elements C, P, and S behave more like high-FIP elements in the SEPs and like low-FIP elements in the SSW. The dashed line at a ratio of 1.2 suggests that O is not the best reference value for the SSW, since it must be slightly elevated by the high crossover for the SSW.

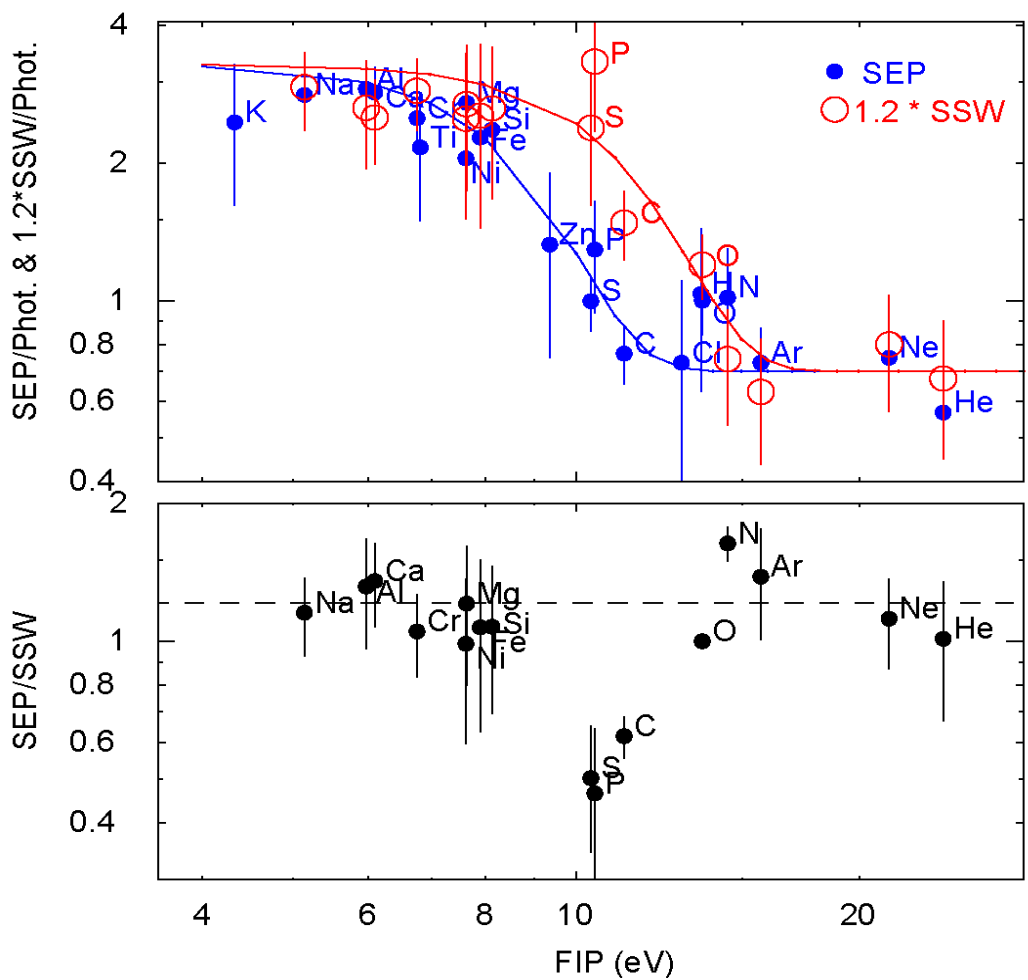

**Figure 8.** The *lower panel* shows the SEP/slow solar wind (SSW) abundance ratio *vs.* FIP for the elements named. The dashed line shows a preferred ratio of ≈1.2. The upper panel shows SEP and 1.2*SSW abundances relative to the photosphere *vs.* FIP with empirical curves to show the trend of the data [13,59]. See Appendix for numerical values.

The first conclusion of this comparison is that the SEPs are not just accelerated solar wind, but rather are an independent and different sample of the solar corona [59]. This has been previously suggested [80,86]. Comparison with theory suggests a possible explanation for the observed difference. In the theory by Laming [71,87], high-FIP neutral atoms propagate across the chromosphere and toward the corona, while the low-FIP ions are also driven upward by the ponderomotive force of Alfvén waves. On closed magnetic loops, the Alfvén wave can resonate with the loop length, on open field lines there can be no such resonance. A lack of resonance changes the ionization pattern of H and makes the intermediate elements C, S, and P behave more like ions. When elements enter the ≈1 MK corona they all become highly ionized. Thus, it has been suggested [59] that comparison with theory shows that particles that would later become SEPs crossed into the corona on closed magnetic loops, while the solar wind crossed on open structures as shown in Figure 9, and recent calculations [88] have improved this comparison. Of course, loops can reconnect and open as plasma evaporates upward, and when shocks accelerate ions to high rigidity, they may no longer be trapped on loops. SEPs are accelerated high in the corona, impulsive SEPs at ≈1.5 $R_s$ [65] and gradual SEPs at 2 – 3 $R_s$ [14,15,89].

As the shock waves in SEP events move out from the Sun, their main input is the SEPs accelerated earlier when that shock was stronger nearer the Sun. However, there are shock waves that do directly accelerate ions from the solar wind. These shocks are produced at corotating interacting regions (CIRs), where high-speed solar-wind streams collide with SSW emitted earlier in the solar rotation. These shocks often form out beyond the orbit of Earth and the accelerated ions show power-law abundance variations in *A/Q* as they flow inward. CIR measurements [90] provide an additional measurement of the FIP-dependence of the solar wind abundances [59,88] and they are included in Figure 9.

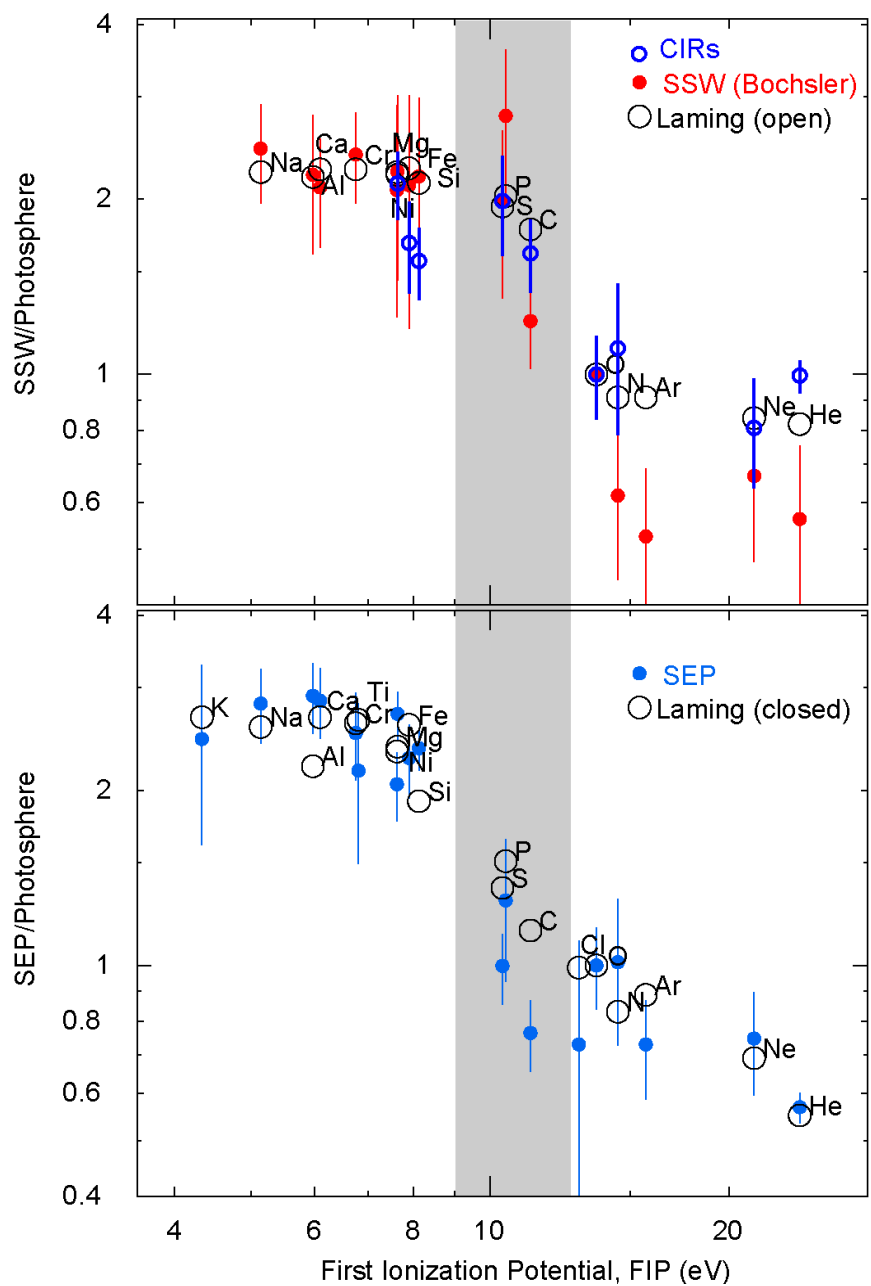

**Figure 9.** The *lower panel* compares the FIP pattern of SEPs [59] with the closed loop model of Laming [87] (Table 3). The *upper panel* compares the SSW [85] and corotating interacting region (CIR) [59] FIP patterns the open field model of Laming [87] (Table 4) [59]. The gray band highlights the region of differences of C, P, and S between the two populations [13,59].

The difference between the FIP patterns of SEPs and the solar wind is important, because such a FIP-dependent difference occurs at the base of the corona, long before the acceleration of either SEPs or the solar wind. It is not terribly surprising that the plasma, which will later become SEPs, originates on closed field lines that occur in or near active regions. However, it is more surprising that *all* of the solar wind plasma originates on open field lines, including the SSW. It is not sufficient that the field lines open later to produce the SSW, they must be open when that plasma crosses the chromosphere and enters the corona, since that is when its FIP is, or is not, controlled by resonant Alfvén waves. The difference that we have found between SEPs and the solar wind is as important for understanding the solar wind as for understanding SEPs [13,59,87,88].

## 5. Conclusions

The element abundances in SEPs are controlled by a complex variety of physical processes, but our understanding of those processes has considerably improved with time. We have come a long way from the birdcage model. There are two major processes involved in the acceleration of the SEPs that we observe in space: magnetic-reconnection in jets and CME-driven shock waves. Resonant wave-particle interactions are important in the enhancements of $^3$He in jets, but, of course, they are also important for the waves that scatter ions back and forth across shocks to produce acceleration. We have discussed four subclasses of the impulsive and gradual SEP events:

- "Pure" impulsive SEPs from jets: Magnetic reconnection in jets produces steep power-law enhancements in *A/Q* from H to Pb and in $^3$He/$^4$He from the ≈3 MK plasma at ≈1.5 $R_s$. Jet ejecta are too slow to produce a shock wave. Flares involve similar physics on *closed* loops, but those energetic ions cannot escape to become SEPs; they deposit their energy in loop footpoints to produce hot, bright flares instead.
- Impulsive SEPs with shocks: An impulsive SEP jet where the CME from that same jet can drive a sufficiently fast shock to favor the reacceleration of the pre-enhanced ≈3 MK impulsive suprathermal ions, but also to sample protons from the ambient plasma to produce a proton excess.
- A weak gradual SEP event: A wide, moderately-fast (>500 km s$^{-1}$) CME-driven shock wave sweeps up protons from the ambient corona, but strongly favors the faster pre-enhanced residual impulsive suprathermal ions that have accumulated in the corona from many small jets. The ≈3 MK impulsive ions dominate at $Z$ >2, but there are excess protons.
- A strong gradual SEP event: A wide, fast (>1000 km s$^{-1}$) CME-driven shock wave accelerates ions predominantly from the ambient 1–2 MK coronal plasma at 2–3 $R_s$. Any impulsive suprathermal ions included are negligible.

The element abundances in the underlying coronal plasma sampled by the SEPs differ from those that were measured in the photosphere by the "FIP effect" shown in Figure 2. The corona sampled by the SEPs differs from the version seen in the solar wind, especially in the intermediate-FIP elements C, P, and S (Figure 8). This difference might occur because the plasma that will later become SEPs enters the corona on closed field lines in active regions, while that producing solar wind arises on open field lines. This distinction helps us to identify the source of the solar wind as well as the source of SEPs.

Apart from known variations in H and He, SEP abundances have a well-defined FIP dependence and smooth dependence on *A/Q*, which is reasonably well understood. Some of the high-FIP abundances, especially C/O, but possibly also Ne/O and Ar/O, lie below the accepted photospheric values; perhaps these abundance in SEPs could be better measures of photospheric abundances of these elements than those that were obtained by other techniques.

**Funding:** This research received no external funding.

**Conflicts of Interest:** The author declares no conflict of interest.

## Appendix

Table 1 shows the atomic number, FIP, and abundances of dominant elements in the solar photosphere as measures by two observers [61,60] and in the solar corona as measured in SEPs [1,52,58,59] and in the slow or interstream solar wind [85]. All abundances are normalized at O = 1000. Elements not modified by Caffau *et al.* [61, indicated by *] retained their ratios to O from Lodders *et al.* [62].

**Table 1.** Element Abundances of the Photosphere and of the Corona as measured by SEPs and the SSW.

|    | Z  | FIP [eV] | Photosphere [*61, 62]        | Photosphere [60]           | SEP [1,52,58]           | SSW [85]      |
|----|----|----------|------------------------------|----------------------------|--------------------------|---------------|
| H  | 1  | 13.6     | $(1.74\pm04)\times10^{6\,*}$ | $(2.04\pm0.05)\times10^6$ | $(1.6\pm0.2)\times10^6$ | –             |
| He | 2  | 24.6     | $(1.46\pm07)\times10^5$      | $(1.74\pm0.04)\times10^5$ | $(0.91\pm03)\times10^5$ | 90000±30000   |
| C  | 6  | 11.3     | 550±76*                      | 550±63                     | 420±9                    | 680±70        |
| N  | 7  | 14.5     | 126±35*                      | 138±16                     | 128±3                    | 78±5          |
| O  | 8  | 13.6     | 1000±161*                    | 1000±115                   | 1000±10                  | 1000          |
| Ne | 10 | 21.6     | 195±45                       | 174±40                     | 157±10                   | 140±30        |
| Na | 11 | 5.1      | 3.47±0.24                    | 3.55±0.33                  | 10.4±1.1                 | 9.0±1.5       |
| Mg | 12 | 7.6      | 60.3±8.3                     | 81.3±7.5                   | 178±4                    | 147±50        |
| Al | 13 | 6.0      | 5.13±0.83                    | 5.75±0.40                  | 15.7±1.6                 | 11.9±3        |
| Si | 14 | 8.2      | 57.5±8.0                     | 66.1±4.6                   | 151±4                    | 140±50        |
| P  | 15 | 10.5     | 0.501±0.046*                 | 0.525±036                  | 0.65±0.17                | 1.4±0.4       |
| S  | 16 | 10.4     | 25.1±2.9*                    | 26.9±1.9                   | 25±2                     | 50±15         |
| Cl | 17 | 13.0     | 0.55±0.38                    | 0.65±0.45                  | 0.24±0.1                 | –             |
| Ar | 18 | 15.8     | 5.5±1.3                      | 5.1±1.5                    | 4.3±0.4                  | 3.1±0.8       |
| K  | 19 | 4.3      | 0.224±0.046*                 | 0.22±0.045                 | 0.55±0.15                | –             |
| Ca | 20 | 6.1      | 3.72±0.60                    | 4.47±0.41                  | 11±1                     | 8.1±1.5       |
| Ti | 22 | 6.8      | 0.138±0.019                  | 0.182±021                  | 0.34±0.1                 | –             |
| Cr | 24 | 6.8      | 0.759±0.017                  | 0.89±0.08                  | 2.1±0.3                  | 2.0±0.3       |
| Fe | 26 | 7.9      | 57.6±8.0*                    | 64.6±6.0                   | 131±6                    | 122±50        |
| Ni | 28 | 7.6      | 2.95±0.27                    | 3.39±0.31                  | 6.4±0.6                  | 6.5±2.5       |
| Zn | 30 | 9.4      | 0.072±0.025                  | 0.074±009                  | 0.11±0.04                | –             |

Note: * meand ref 61